# Collisionless amplifying of longitudinal electron waves in two-stream plasma[1]

V. N. Soshnikov [2]
Plasma Physics Dept.,
All-Russian Institute of Scientific and Technical Information
of the Russian Academy of Sciences
*(VINITI, Usievitcha 20, 125315 Moscow, Russia)*

*Abstract*

*To better understanding the principal features of collisionless damping/growing plasma waves we have implemented a demonstrative calculation for the simplest cases of electron waves in two-stream plasmas with the δ-type electron velocity distribution function of each of the streams with velocities $v_1$ and $v_2$. The traditional dispersion equation is reduced to an algebraic 4th order equation, for which numerical solutions are presented for a variant of equal stream densities. In the case of a uniform half-infinite slab one finds two dominant type solutions: non-damping forward waves and forward complex conjugated exponentially damping and growing waves. Beside it in this case there is no necessity of calculation any logarithmically divergent uncertain integrals. The possibility of wave amplifying might be useful in practical applications.*

**Introduction**

This paper is continuation of the before presented works [1 – 7] with consideration of the relatively simplest problem of electron waves in half-infinite slab of homogeneous collisionless plasma with special stress on the principal aim of some better understanding fundamental laws of wave solutions.

In the end, this problem reduces to finding precise or approximate solutions of the traditional dispersion equation applying Laplace transform method and to non-traditional analysis of corresponding them asymptotical solutions of wave equations.

As a boundary condition one takes usually the condition

$$E(x=0) = E_0 \exp(i\omega t), \qquad (1)$$

where $\omega$ is frequency of applied boundary electrical field exciting plasma oscillations and waves, $x$ is coordinate along direction of wave propagation.

The boundary electron velocity distribution function is then not arbitrary and must be found according to equations with a series of additional physical conditions. The corresponding asymptotical solution at large $(x,t)$ is

$$E = \sum_l a_l \exp(i\omega t - ik_l x), \qquad (2)$$

---

[1] It is English version of the paper published in Russian "The Integrated Scientific Journal", 2006, n.4, p. 61.
[2] Krasnodarskaya str., 51-2-168, Moscow 109559, Russia. E-mail: vikt3363@yandex.ru .

with summing over all residua in Laplace transform method. It appears evident that transition to practically measurable real number values is fulfilled by taking real number values of boundary conditions and of corresponding asymptotical values (that is equivalent to summing conjugated solutions at boundary conditions $E = E_0 e^{i\omega t}/2$ and $E = E_0 e^{-i\omega t}/2$ ),

$$E(x=0) = E_0 \cos \omega t = \text{Re}\left[ E_0 \exp(i\omega t) \right], \qquad (3)$$

what corresponds to the solution due to the boundary field

$$E_0 \cos(\omega t) = E_0 \cos(-\omega t). \qquad (4)$$

Expression (4) together with functions $k(\omega)$ for the different roots $k_l$ of dispersion equation determines the type of solution: either only forward waves, par example at $(\omega, k_l) > 0$ and $k_l(-\omega) = -k_l(\omega)$ at $\omega < 0$, or simultaneous presence of forward waves with exponentially growing and damping amplitudes at complex number $k_l$, and also a possibility of presence of backward damping and growing waves.

We consider further the simplest case of waves in two-stream plasma with the distribution function

$$f(\vec{v}) = a\delta(v_x - v_1)\delta(v_y)\delta(v_z) + b\delta(v_x - v_2)\delta(v_y)\delta(v_z), \qquad (5)$$

where $f(\vec{v})$ is normalized to unity and

$$a + b = 1. \qquad (6)$$

One considers only longitudinal waves which are solutions of dispersion equation

$$1 + \frac{\omega_L^2}{k} \int \frac{\partial f/\partial v_x}{\omega - kv_x} dv_x = 0, \qquad (7)$$

where $\omega_L$ is Langmuir frequency, and $k$ is wave number. Quadratic non-linear term of wave equations is neglected, so far as its account for leads only to appearance of overtones with multiple frequencies at slight influence upon the waves at the given boundary frequency [2].

**Demonstrative solutions of dispersion equation**

In all analogy with [6] where one considered one-stream plasma with $\delta$-like electron velocity distribution function, dispersion equation for two-stream distribution function (5), (6) after integrating by parts will be

$$1 = \frac{a\omega_L^2}{(\omega - kv_1)^2} + \frac{b\omega_L^2}{(\omega - kv_2)^2}, \qquad (8)$$

where $v_1$, $v_2$ are stream velocities in direction $x$.

Defining

$$Z \equiv \frac{kv_2}{\omega}; \quad \beta \equiv \frac{v_1}{v_2}; \quad v_1, v_2 > 0, \qquad (9)$$

we obtain an fourth-order equation for finding $Z$:

$$\beta^2 Z^4 - 2\beta(1+\beta)Z^3 + AZ^2 + BZ + \left(1 - \frac{\omega_L^2}{\omega^2}\right) = 0, \qquad (10)$$

where

$$A \equiv 1 + 4\beta + \beta^2 - \frac{\omega_L^2}{\omega^2}\left(a + \beta^2 - a\beta^2\right), \qquad (11)$$

$$B \equiv 2\frac{\omega_L^2}{\omega^2}(a + \beta - a\beta) - 2(1+\beta). \qquad (12)$$

As an illustration of some characteristic features of solutions of Eq.10 we present results of number solution of an arbitrary selected variant with

$$a = b = 0.5 \qquad (13)$$

at various values $\omega_L^2/\omega^2$ and $\beta \equiv v_1/v_2$ (see Table 1).

**Discussion**

For the most part the calculated solutions are positive; there are also complex conjugated roots corresponding to exponentially growing and damping waves. Since the wave speed is

$$v_{wave} = \frac{\omega}{\operatorname{Re} k} = \frac{v_2}{\operatorname{Re} Z} = \frac{v_1/\beta}{\operatorname{Re} Z}, \qquad (14)$$

and $Z$ is independent of the frequency sign $\pm\omega$, values $\operatorname{Re} Z > 0$ correspond only to forward waves, in this case there are no backward waves with the same frequency $|\omega|$ and $\omega/\operatorname{Re} k < 0,$ so as at $\operatorname{Re} Z < 0$ analogously there would be no forward waves with the same frequency $|\omega|$. Therefore one can assume that at absence of backward waves which could excite plasma turbulence both damping and growing modes of forward waves are really existing. The same assertion is justified also for group waves. Defining $\eta \equiv \omega_L^2/\omega^2$ and accounting for (9) one obtains

$$v_{group} \equiv \frac{d\omega}{d\operatorname{Re} k} = \frac{v_2}{\operatorname{Re} Z - 2\dfrac{\eta d\operatorname{Re} Z}{d\eta}} \simeq \frac{v_2}{\operatorname{Re} Z} \qquad (15)$$

due to smallness of $\left|\dfrac{\eta d\operatorname{Re} Z}{d\eta}\right|$ following from the Table 1.

Returning to Maxwellian distribution function in preceding papers and Landau rule of passing around poles, it ought to note that due to quadratic character of approximate dispersion equation (see [3]) solution is determined by the principal sense value of the logarithmically divergent integral and has the form of simultaneously existent both forward and backward waves (at the boundary condition (1)). At account for small complex number additions at taking the logarithmically divergent integral according to Landau rule we obtain both forward wave, exponentially damping at growing $x$, and backward wave exponentially growing at growing $x$, or, according to the direction of passing around poles, forward wave growing at growing $x$, and backward wave damping at growing $x$. In both cases appearing turbulence might be transported by the background wave to small values of coordinate $x$ with arising an instability at all values $x$.

*Table* **1**

*Roots $Z = kv_2/\omega$ of the dispersion equation for longitudinal waves in two-stream plasma for the case of a=b=0,5\**

| $\omega_L^2/\omega^2$ \ $v_1/v_2$ | 0.8 | 0.5 | 0.3 | 0.1 | 0.05 | 0 |
|---|---|---|---|---|---|---|
| 0.8 | 1.110±0.1287$i$<br>1.140±0.923$i$ | 2.59<br>1.57<br>0.920±0.657$i$ | 5.85<br>1.595<br>0.611±0.777$i$ | 21.04<br>1.717<br>–0.378±0.596$i$ | 43.70<br>1.762<br>0.0075<br>–3.47 | 1.816<br>0.184 |
| 0.5 | 1.121±0.1333$i$<br>1.129±0.683$i$ | 2.97<br>1.388<br>0.822±0.544$i$ | 6.17<br>1.443<br>0.528±0.609$i$ | 21.70<br>1.525<br>0.246<br>−1.472 | 44.94<br>1.550<br>0.361<br>−4.85 | 1.577<br>0.423 |
| 0.3 | 1.152±0.1403$i$<br>1.098±0.458$i$ | 3.16<br>1.290<br>0.773±0.433$i$ | 6.36<br>1.338<br>0.483±0.429$i$ | 22.12<br>1.391<br>0.513<br>−2.03 | 45.73<br>1.405<br>0.553<br>−5.69 | 1.420<br>0.580 |
| 0.1 | 1.443<br>1.137<br>0.960±0.223$i$ | 3.34<br>1.173<br>0.746±0.223$i$ | 6.55<br>1.199<br>0.669<br>0.254 | 22.53<br>1.220<br>0.756<br>−2.50 | 46.50<br>1.225<br>0.763<br>−6.48 | 1.229<br>0.771 |
| 0.05 | 1.531<br>1.099<br>1.064±0.1622$i$ | 3.38<br>1.129<br>0.479±0.1344$i$ | 6.59<br>1.144<br>0.805<br>0.1281 | 22.63<br>1.155<br>0.833<br>−2.62 | 46.68<br>1.161<br>0.834<br>−6.68 | 1.160<br>0.84 |
| 0 | 1.25<br>1 | 2.00<br>1 | 3.33<br>1 | 10.0<br>1 | ∞<br>1 | –<br>1 |

\* In every column of the table the roots are written in a sequence, generally speaking not correlated with the root sequences in other columns.

Applying of Laplace method leads to a multitude of mathematically correct solutions of dispersion equation (in violation of the uniqueness theorem) depending on selected way of taking indefinitely divergent integral in dispersion equation, particularly at applying Landau rule for contour integral.

One of physical requirements in the case of half-infinite slab appears to be absence of the so-called kinematical waves not connected with the given boundary electrical field and also absence (in rest plasma) of unphysical backward waves.

The requirement of absence of kinematical waves leads to an integral equation [3,4] determining boundary distribution function which according to Landau rule must contain complex value addition to amplitude depending on $v_x$ at the common for all constituent terms factor $\exp(i\omega t)$, what does not agree with the real value amplitude of exciting boundary field. Thus, there is no solution of equations which would satisfy both requirements [3]. In this case exponentially growing solutions are unavoidable at any choice of boundary conditions for distribution function.

In is surprising that in the considered here two-stream plasma problem indefinitely divergent integrals are absent, correspondingly any difficulties of their calculation with using Landau rule do not arise.

Beside the foregoing consideration of longitudinal waves we have carried out analogous calculations for transversal waves in two-stream plasma [7]. In this case dispersion equation reduces to

$$-k^2 + \frac{\omega^2}{c^2} - \frac{\omega_L^2 \omega}{c^2}\left(\frac{a}{\omega - kv_1} + \frac{b}{\omega - kv_2}\right) = 0; \quad a + b = 1. \quad (16)$$

In difference of (8) this equation includes an additional parameter $\gamma \equiv v_2^2/c^2$.

As an illustration of general features of solutions we present some results of number solving fourth-order Eq. 16 at various $\omega_L^2/\omega^2$, $\beta = v_1/v_2$ and arbitrarily selected values $a = b = 0.5$; $\gamma = 3.91 \times 10^{-3}$, the latter corresponds to electron energy $m_e v_2^2/2 = 1$ keV. The solution has the form of simultaneously existent fast and slow wave modes, the unphysical fast backward wave can be removed with the corresponding choice of lacking boundary conditions. After that every solution reduces to only three forward waves. The absence in this solutions of amplitude damping/growing appears to be very surprising (see Table 2). The neglected qualitative term of kinetic equation does not change qualitative the character of solutions leading only to appearance of relatively weak overtones with multiple frequencies [2], [3], [7].

There is open question about finding precise solutions in the region of saturation $x > x_0$ where distribution function could become negative. It is evidently that in this region the traditional kinetic equation becomes inapplicable. One assumed however that in this region it ought to put the distribution function equal zero as it was proposed in [2], [3].

*Table* **2**

Roots $Z = kv_2/\omega$ of dispersion equation of transverse waves in two-stream plasma for variant $a = b = 0.5$; $\gamma = v_2^2/c^2 = 3.91 \times 10^{-3}$  *

| $\eta = \omega_L^2/\omega^2$ \ $v_1/v_2$ | 0.8 | 0.6 | 0.4 | 0.2 |
|---|---|---|---|---|
| 0.8 | -0.0294<br>0.0266<br>1.002<br>1.251 | -0.0292<br>0.0267<br>1.002<br>1.668 | -0.0291<br>0.0269<br>1.002<br>2.50 | -0.0309<br>0.0251<br>1.006<br>5.00 |
| 0.6 | -0.0406<br>0.0385<br>1.001<br>1.251 | -0.0405<br>0.0386<br>1.001<br>1.667 | -0.0404<br>0.0387<br>1.001<br>2.50 | -0.0402<br>0.0388<br>1.001<br>5.00 |
| 0.4 | -0.0491<br>0.0477<br>1.001<br>1.251 | -0.0490<br>0.0478<br>1.001<br>1.667 | -0.0490<br>0.0479<br>1.001<br>2.50 | -0.0439<br>0.0479<br>1.001<br>5.00 |
| 0.2 | -0.0563<br>0.0556<br>1.000<br>1.250 | -0.0562<br>0.0556<br>1.000<br>1.667 | -0.0562<br>0.0556<br>1.000<br>2.50 | -0.0562<br>0.0557<br>1.000<br>5.00 |

* Group velocity of waves is determined with expression
$$v_{group} = \frac{0{,}0625c}{\operatorname{Re} Z - 2\eta \dfrac{\partial \operatorname{Re} Z}{\partial \eta}}$$

**Conclusion**

We have found demonstrative four-modes solutions of traditional dispersion equation in the case of longitudinal and transverse electron waves in half-infinite slab of collisionless two-stream plasma. Both parameter regions are present, as with four non-damping forward wave modes as well as instability regions with two complex conjugated forward wave modes corresponding to both exponentially damping and growing waves.

In the considered cases dispersion equations do not contain indefinitely divergent integrals, therefore no necessity is to use any procedures of calculation of such logarithmically divergent integrals as well as Landau rule of passing around poles in the complex value velocity plane.

Once again it ought to note that in the case of rest plasma with Maxwellian velocity distribution function the application of Landau rule leads to impossibility of satisfying the physical requirements of absence of kinematical and backward exponentially either growing or damping waves independently of passing around poles.

The presence of only forward exponentially growing waves at possibility of removing backward waves imply practical chance of using effect of wave amplifying in two-stream plasmas.